\RequirePackage{amsmath}
\documentclass[11pt,onecolumn,a4paper]{article}
\usepackage[utf8]{inputenc}
\usepackage[english]{babel}
\usepackage[latexsym]{}
\usepackage{amsmath}
\usepackage{amsfonts} 

\usepackage{graphicx}
\usepackage{color}
\usepackage{soul}
\usepackage{enumitem}
\usepackage{hyperref}
\usepackage{url,footnote}
\usepackage{stmaryrd}
\usepackage[T1]{fontenc}
\usepackage{sidecap}
\usepackage{wrapfig}
\usepackage{cleveref}
\usepackage{multirow}
\usepackage{listings}
\usepackage{amssymb}
\usepackage[linesnumbered,ruled,lined,boxed,commentsnumbered]{algorithm2e}
\usepackage{ifsym}
\usepackage{stmaryrd}
\usepackage{booktabs}
\usepackage{flushend}
\usepackage{balance}

\newcommand{\code}[1]{\texttt{\small#1}} 
    
\newcommand{\powerset}{\mathcal{P}}

\newcommand{\dom}{\operatorname{dom}}
\newcommand{\range}{\operatorname{range}}
\newcommand{\type}{\operatorname{type}}
\newcommand{\sub}{\operatorname{sub}}
\newcommand{\pred}{\operatorname{pred}}
\newcommand{\obje}{\operatorname{obj}}
\newcommand{\prefix}{\operatorname{prefix}}
\newcommand{\name}{\operatorname{name}}

\newcommand{\dI}{\textbf{I}}
\newcommand{\dL}{\textbf{L}}

\newcommand{\dLab}{\mathbb{L}}
\newcommand{\dV}{\mathbb{V}}
\newcommand{\dT}{\mathbb{T}}
\newcommand{\dN}{\mathbb{N}}
\newcommand{\dE}{\mathbb{E}}
\newcommand{\dP}{\mathbb{P}}

\newcommand{\sm}{\mathcal{SM}}
\newcommand{\im}{\mathcal{IM}}
\newcommand{\dm}{\mathcal{DM}}

\newtheorem{theorem}{Theorem}
\newtheorem{lemma}{Lemma}
\newtheorem{corollary}{Corollary}
\newtheorem{proposition}{Proposition}
\newtheorem{example}{Example}[section]
\newtheorem{definition}{Definition} 

\definecolor{codegreen}{rgb}{0,0.6,0}
\definecolor{codegray}{rgb}{0.5,0.5,0.5}
\definecolor{codepurple}{rgb}{0.58,0,0.82}
\definecolor{backcolour}{rgb}{0.95,0.95,0.92}
 
\lstdefinestyle{mystyle}{
    backgroundcolor=\color{backcolour},   
    commentstyle=\color{codegreen},
    keywordstyle=\color{magenta},
    numberstyle=\tiny\color{codegray},
    stringstyle=\color{codepurple},
    basicstyle=\normalsize,
    breakatwhitespace=false,         
    breaklines=true,                 
    captionpos=b,                    
    keepspaces=true,                 
    numbers=left,                    
    numbersep=5pt,                  
    showspaces=false,                
    showstringspaces=false,
    showtabs=false,                  
    tabsize=2,
    mathescape
}
 
\hypersetup{
    colorlinks=true,
    linkcolor=blue,
    filecolor=red,      
    urlcolor=cyan,
}

\title{Directly Mapping RDF Databases to Property Graph Databases\footnote{A  newer version of this article has been accepted and published at the IEEE Access Journal, DOI: \texttt{\url{10.1109/ACCESS.2020.2993117}}. Please refer to and cite~\cite{9088985} for the latest version of this article.}}

\author{Renzo Angles, Harsh Thakkar, Dominik Tomaszuk}

\date{}

\begin{document}

\maketitle

\begin{abstract}
RDF triplestores and property graph databases are two approaches for data management which are based on modeling, storing and querying graph-like data. In spite of such common principle, they present special features that complicate the task of database interoperability. While there exist some methods to transform RDF graphs into property graphs, and vice versa, they lack compatibility and a solid formal foundation. This paper presents three direct mappings (schema-dependent and schema-independent) for transforming an RDF database into a property graph database, including data and schema. We show that two of the proposed mappings satisfy the properties of semantics preservation and information preservation. The existence of both mappings allows us to conclude that the property graph data model subsumes the information capacity of the RDF data model. 
\end{abstract}

\section{Introduction}
RDF~\cite{10008} and Graph databases~\cite{70101} are two approaches for data management that are based on modeling, storing and querying graph-like data.
The database systems based on these models are gaining relevance in the industry due to their use in various application domains where complex data analytics is required~\cite{91531}.

RDF triplestores and graph database systems are tightly connected as they are based on graph data models.
RDF databases are based on the RDF data model~\cite{10008}, their standard query language is SPARQL~\cite{sparql11}, and RDF Schema~\cite{rdfschema} allows to describe classes of resources and properties (i.e. the data schema). 
On the other hand, most graph databases are based on the Property Graph (PG) data model, there is no standard query language, and there is no standard notion of property graph schema~\cite{91372}. 
Therefore, RDF and PG database systems are dissimilar in data model, schema constraints and query language.


\smallskip
\noindent
\textbf{Motivation.}
The term ``Interoperability'' was introduced in the area of information systems, and is defined as the ability of two or more systems or components to exchange information, and to use the information that has been exchanged \cite{91491}.
In the context of data management, interoperability is concerned with the support of applications that share and exchange information across the boundaries of existing databases \cite{91396}.

Databases interoperability is relevant for several reasons, including: promotes data exchange and data integration \cite{91389,nguyen2019singleton}; facilitates the reuse of available systems and tools \cite{DBLP:journals/corr/abs-1910-03118,91396}; enables a fair comparison of database systems by using benchmarks \cite{91027,DBLP:conf/esws/Thakkar17,DBLP:conf/i-semantics/ThakkarKDLA17}; and supports the success of emergent systems and technologies \cite{91396,thakkar2020let}.

Given the heterogeneity between RDF triplestores and graph database systems, and considering their graph-based data models, it becomes necessary to develop methods to allow interoperability among these systems. 

\smallskip
\noindent
\textbf{The Problem.}
To the best of our knowledge, the research about the interoperability between RDF and PG databases is very restricted (cf. Section~\ref{sec:relwork}). 
While there exist some system-specific approaches, most of them are restricted to data transformation and lack of solid formal foundations. 

\smallskip
\noindent
\textbf{Objectives \& Contributions.}
Database interoperability can be divided into syntactic interoperability (i.e. data format transformation), semantic interoperability (i.e. data exchange via schema and instance mappings) and query interoperability (i.e. transformations among different query languages or data accessing methods) \cite{AnglesAMW19}.

The main objective of this paper is to study the semantic interoperability between RDF and PG databases. Specifically, we propose two mappings to translate RDF databases into PG databases. We study two desirable properties of these database mappings, named semantics preservation and information preservation. Based on such database mappings, we conclude that the PG data model subsumes the information capacity of the RDF data model.

The remainder of this paper is as follows: 
A formal background is presented in Section~\ref{sec:gdm}, including  definitions related to database mappings, RDF databases, and Property Graph databases;
A schema-dependent database mapping, to transform RDF databases into PG databases, is presented in Section~\ref{sec:dm1};
A schema-independent database mapping is presented in Section~\ref{sec:dm2};
The related work is presented in Section~\ref{sec:relwork};
Our conclusions are presented in Section~\ref{sec:conclu}.

\section{Preliminaries}
\label{sec:gdm}
This section presents a formal background to study the interoperability between RDF and PG databases. In particular, we formalize the notions of database mapping, RDF database, and property graph database.

\subsection{Database mappings}
\label{sec:maps}
In general terms, a database mapping is a method to translate databases from a source database model to a target database model.
We can consider two types of database mappings:
\emph{direct database mappings}, which allow an automatic translation of databases without any input from the user \cite{91025}; 
and \emph{manual database mappings}, which require additional information (e.g. an ontology) to conduct the database translation.
In this paper, we focus on direct database mappings.

\paragraph{Database schema and instance}
Let $M$ be a database model.
A \emph{database schema} in $M$ is a set of semantic constraints allowed by $M$.
A \emph{database instance} in $M$ is a collection of data represented according to $M$. 
A \emph{database} in $M$ is an ordered pair $D^M = (S^M,I^M)$, where $S^M$ is a schema and $I^M$ is an instance.

Note that the above definition does not establish that the database instance satisfies the constraints defined by the database schema. 
Given a database instance $I^M$ and a database schema $S^M$, we say that $I^M$ is valid with respect to $S^M$, denoted $I^M \models S^M$, iff $I^M$ satisfies the constraints defined by $S^M$.
Given a database $D^M = (S^M,I^M)$, we say that $D^M$ is a valid database iff it satisfies that $I^M \models S^M$.

\paragraph{Schema, instance, and database mapping}
A database mapping defines a way to translate databases from a ``target'' database model to a ``source'' database model. For the rest of this section, assume that $M_1$ and $M_2$ are the source and the target database models respectively.

Considering that a database includes a schema and an instance, we first define the notions of schema mapping and instance mapping.
A \emph{schema mapping} from $M_1$ to $M_2$ is a function $\sm$ from the set of all database schemas in $M_1$, to the set of all database schemas in $M_2$. 
Similarly, an \emph{instance mapping} from $M_1$ to $M_2$ is a function 
$\im$ from the set of all database instances in $M_1$, to the set of all database instances in $M_2$.

A database mapping from $M_1$ to $M_2$ is a function $\dm$ from the set of all databases in $M_1$, to the set of all databases in $M_2$. Specifically, a database mapping is defined as the combination of a schema mapping and an instance mapping.  

\smallskip
\begin{definition}[Database Mapping]
A database mapping is a pair $\dm = (\sm,\im)$ 
where $\sm$ is a schema mapping and $\im$ is an instance mapping.
\end{definition}
\smallskip

\subsubsection{Properties of database mappings}
\label{sec:properties}
Every data model allows to structure the data in a specific way, or using a particular abstraction. 
Such abstraction determines the conceptual elements that the data model can represent, i.e. its representation power or information capacity \cite{50604}.

Given two database models $M_1$ and $M_2$, the possibility to exchange databases between them depends on their information capacity. 
Specifically, we say that $M_1$ subsumes the information capacity of $M_2$ iff every database in $M_2$ can be translated to a database in $M_2$. 
Additionally, we say that $M_1$ and $M_2$ have the same information capacity iff $M_1$ subsumes $M_2$ and $M_2$ subsumes $M_1$.
 
The information capacity of two database models can be evaluated in terms of a database mapping satisfying some properties. In particular, we consider three properties: computability, semantics preservation, and information preservation.

Assume that $\mathcal{D}_1$ is the set of all databases in a source database model $M_1$, and $\mathcal{D}_2$ is the set of all databases in a target database model $M_2$. 

\smallskip
\begin{definition}[Computable mapping]
A database mapping $\dm : \mathcal{D}_1 \to \mathcal{D}_2$ is computable if there exists an algorithm $\mathcal{A}$ that, given a database $D \in \mathcal{D}_1$, $\mathcal{A}$ computes $\dm(D)$.
\end{definition}
\smallskip

The property of computability indicates the existence and feasibility of implementing a database mapping from $M_1$ to $M_2$. This property also implicates that $M_2$ subsumes the information capacity of $M_1$. 

\smallskip
\begin{definition}[Semantics preservation]
A computable database mapping $\dm : \mathcal{D}_1 \to \mathcal{D}_2$ is semantics preserving if for every valid database $D \in \mathcal{D}_1$, there is a valid database $D' \in \mathcal{D}_2$ satisfying that $D' = \dm(D)$.   
\end{definition}
\smallskip

Semantics preservation indicates that the output of a database mapping is always a valid database. Specifically, the output database instance satisfies the constraints defined by the output database schema. In this sense, we can say that this property evaluates the correctness of a database mapping.

\smallskip
\begin{definition}[Information preservation]
A database mapping $\dm = (\sm, \im)$ from $M_1$ to $M_2$ is information preserving if there is a computable database mapping 
$\dm^{-1} = (\sm^{-1}, \im^{-1})$ from $M_2$ to $M_1$ such that for every database $D = (S,I)$ in $M_1$, it applies that $D = \dm^{-1}( \dm(D) )$.
\end{definition} 
\smallskip

Information preservation indicates that, for some database mapping $\dm_1$, there exists an ``inverse'' database mapping $\dm^{-1}$ which allows to recover a database transformed with $\dm_1$.
Note that the above definition implies the existence of both a ``inverse'' schema mapping $\sm^{-1}$ and a ``inverse'' instance mapping $\im^{-1}$.  
  
Information preservation is a fundamental property because it guarantees that a database mapping does not lose information \cite{91025}.    
Moreover, it implies that the information capacity of the target database model subsumes the information capacity of the source database model.

Our goal is to define database mappings between the RDF data model and the Property Graph data model.
Hence, next, we will present a formal definition of the notions of instance, schema, and database for them.  

\subsection{RDF Databases}
\label{sec:rdf_data_model}
An RDF database is an approach for data management which is oriented to describe the information about Web resources by using Web models and languages. In this section we describe two fundamental standards used by RDF databases: the Resource Description Framework (RDF) \cite{10008}, which is the standard data model to describe the data; and RDF Schema \cite{rdfschema}, which is a standard vocabulary to describe the structure of the data.

\subsubsection{RDF Graph.}
Assume that $\dI$ and $\dL$ are two disjoint infinite sets, called IRIs and Literals respectively. IRIs are used as web resource identifiers and Literals are used as values (e.g. strings, numbers or dates). 
In addition to IRIs and Literals, the RDF data model considers a domain of anonymous resources called Blank Nodes. Based on the work of Hogan et al. \cite{91033}, we avoid the use of Blank Nodes as that their absence does not affect the results presented in this paper. Moreover, we can obtain similar results by replacing Blank Nodes with IRIs (via Skolemisation~\cite{91033}).

An RDF triple is a tuple $t = (v_1,v_2,v_3)$ where $v_1 \in \dI$ is called the \emph{subject}, $v_2 \in \dI$ is the \emph{predicate} and $v_3 \in \dI \cup \dL$ is the \emph{object}.
Here, the subject represents a resource (identified by an IRI), the predicate represents a relationship of the resource (identified by an IRI), and the object represents the value of such relationship (which is either an IRI or a literal).

Let $S$ be a set of RDF triples. We use $\sub(S)$, $\pred(S)$ and $\obje(S)$ to denote the sets of subjects, predicates, and objects in $S$ respectively.    
There are different data formats to encode a set of RDF triples.
The following example shows an RDF description encoded using the Turtle data format \cite{10121}.

\smallskip
\begin{example}
\label{ex:rdfg1}
\begin{lstlisting}
@prefix rdf: <http://www.w3.org/1999/02/22-rdf-syntax-ns#> .
@prefix xsd: <http://www.w3.org/2001/XMLSchema#>
@prefix voc: <http://www.example.org/voc/> .
@prefix ex: <http://www.example.org/data/> .
ex:Tesla_Inc rdf:type voc:Organisation ;
             voc:name "Tesla, Inc." ;
             voc:creation "2003-07-01"^^xsd:date .
ex:Elon_Musk rdf:type voc:Person ;
             voc:birthName "Elon Musk" ;
             voc:age "46"^^xsd:int ;
             voc:ceo ex:Tesla_Inc .
\end{lstlisting}
\end{example}
\smallskip

The lines beginning with \code{@prefix} are prefix definitions and the rest are RDF triples.
A prefix definition associates a prefix (e.g. \code{voc}) with an IRI (e.g. \code{http://www.example.org/voc/}).
Hence, a full IRI like \\ \code{http://www.example.org/voc/Person} can be abbreviated as a prefixed named \code{voc:Person}.
We will use $\prefix(r)$ and $\name(r)$ to extract the prefix and the name of an IRI $r$ respectively. 
We will consider two types of literals:
a literal which consists of a string and a datatype IRI (e.g. \code{"46"\^{}\^{}xsd:int}), and 
a literal which is a Unicode string (e.g. \code{"Elon Musk"}), which is a synonym for a \code{xsd:string} literal (e.g. \code{"Elon Musk"\^{}\^{}xsd:string}).

A set of RDF triples can be visualized as a graph where the nodes represent the resources, and the edges represent properties and values. 
However, the RDF model has a special feature: an IRI can be used as an object and predicate in an RDF graph. For instance, the triple  (\code{voc:ceo}, \code{rdfs:label}, \code{"Chief Executive Officer"}) can be added to the graph shown in Example~\ref{ex:rdfg1} to include metadata about the property \code{voc:ceo}. It implies that an RDF graph is not a traditional graph because it allows edges between edges, and consequently an RDF graph cannot be visualized in a traditional way. Next, we present a formal definition of the RDF data model which supports a traditional graph-based representation.  

\smallskip
\begin{definition}[RDF Graph]\label{def:rdf_graph}
An RDF graph is defined as a tuple 
$G^R = (N_I, N_L, E_O, E_D, \alpha_I, \alpha_L, \beta_O, \beta_D, \delta)$ where:
\begin{itemize}
\item $N_I$ is a finite set of nodes representing RDF resources (i.e. resource nodes);
\item $N_L$ is a finite set of nodes representing RDF literals (i.e. literal nodes), satisfying that $N_I \cap N_L = \emptyset$;
\item $E_O$ is a finite set of edges called object property edges;
\item $E_D$ is a finite set of edges called datatype property edges, satisfying that $E_O \cap E_D = \emptyset$\footnote{We borrowed the names from Web Ontology Language.};
\item $\alpha_I : N_I \rightarrow \dI$ is a total one-to-one function that associates each resource node with a single IRI;
\item $\alpha_L : N_L \rightarrow \dL$ is a total one-to-one function that associates each literal node with a single literal;
\item $\beta_O : E_O \rightarrow (N_I \times N_I)$ is a total function that associates each object property edge with a pair of resource nodes;
\item $\beta_D : E_D \rightarrow (N_I \times N_L)$ is a total function that associates each datatype property edge with a resource node and a literal node;
\item $\delta$: $(N_I \cup N_L \cup E_O \cup E_D) \rightarrow \dI$ is a partial function that assigns a resource class label to each node or edge. 
\end{itemize}
\end{definition}
\smallskip

Note that the function $\delta$ has been defined as being partial to support a partial connection between schema and data (which is usual in real RDF datasets).
Concerning the issue about an IRI $u$ occurring as both resource and property, note that $u$ will occur as resource and property separately. In such a case, we will have a bipartite graph. 

In order to facilitate the transformation of RDF data to Property Graphs, we will assume that every node or edge in an RDF graph defines a resource class. 
This assumption is shown by the following procedure which allows transforming a set of RDF triples into a formal RDF graph.

The procedure to create an RDF Graph from a set $S$ of RDF triples is defined as follows:
\begin{itemize}
\item For every resource $r \in \sub(S)$, there is a node $n \in N_I$ with $\alpha_I(n) = r$;
 \begin{itemize}
  \item If $(r,\code{rdf:type},c) \in S$ 
        then $\delta(n) = c$, 
        else $\delta(n) = \code{rdfs:Resource}$; 
 \end{itemize}
\item For every literal $l \in \obje(S) \cap \dL$, there is a node $n \in N_L$;
 \begin{itemize}
  \item If $l$ is a literal of the form \code{value} then 
        $\alpha_L(n) = l$ and $\delta(n) = \code{xsd:string}$;
  \item If $l$ is a literal of the form \code{value\^{}\^{}datatype} then 
        $\alpha_L(n)$ = \code{value} and $\delta(n)$ = \code{datatype}; 
 \end{itemize}
\item For every triple $(s,p,o) \in S$ where $o \in \dI$, there is an edge $e \in E_O$ with $\delta(e) = p$ and $\beta_O(e) = (n,n')$, such that $\alpha_I(n) = s$ and $\alpha_I(n') = o$;
\item For every triple $(s,p,o) \in S$ where $o \in \dL$, there is an edge $e \in E_D$ with $\delta(e) = p$ and $\beta_D(e) = (n,n')$, such that $\alpha_I(n) = s$ and $\alpha_L(n') = o$.   
\end{itemize}
 
Hence, the RDF graph obtained from the set of RDF triples shown in Example~\ref{ex:rdfg1} is given as follows:

\smallskip
\begin{lstlisting}
$N_I = \{ n_1, n_2 \}$, $N_L = \{ n_3, n_4, n_5, n_6 \}$, 
$E_O = \{ e_1 \}$, $E_D = \{ e_2, e_3, e_4, e_5 \}$,
$\alpha_I(n_1) = \text{ex:tesla\_Inc}$, $\alpha_I(n_2) = \text{ex:Elon\_Musk}$,
$\alpha_L(n_3) = \text{"Tesla, Inc."}$, $\alpha_L(n_4) = \text{"2003-07-01"}$, $\alpha_L(n_5) = \text{"Elon Musk"}$, $\alpha_L(n_6) = \text{46}$,
$\beta_O(e_1) = (n_1, n_2)$,
$\beta_D(e_2) = (n_1, n_3)$, $\beta_D(e_3) = (n_1, n_4)$, $\beta_D(e_4) = (n_2, n_5)$, $\beta_D(e_5) = (n_2, n_6)$,
$\delta(n_1) = \text{voc:Organisation}$, $\delta(n_2) = \text{voc:Person}$, $\delta(n_3) = \text{xsd:string}$, $\delta(n_4) = \text{xsd:date}$, $\delta(n_5) = \text{xsd:string}$, $\delta(n_6) = \text{xsd:int}$,
$\delta(e_1) = \text{voc:ceo}$, $\delta(e_2) = \text{voc:name}$, $\delta(e_3) = \text{voc:creation}$, $\delta(e_4) = \text{voc:birthName}$, $\delta(e_5) = \text{voc:age}$. 
\end{lstlisting}
\smallskip

\begin{figure}[ht]
    \centering
    \includegraphics[width=\linewidth]{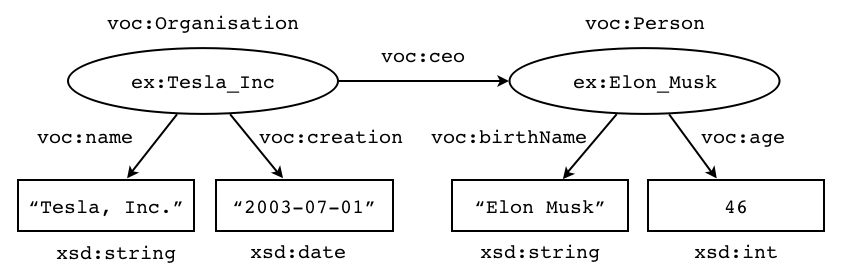}
    \caption{Graphical illustration of an RDF graph.}
    \label{fig:rdfg1}
\end{figure}

Additionally, Figure~\ref{fig:rdfg1} shows a graphical representation of the RDF graph described above.
The resource nodes are represented as ellipses and literal nodes are presented as rectangles.
Each node is labeled with two IRIs: the inner IRI indicates the resource identifier, and the outer IRI indicates a resource class. 
Each edge is labeled with an IRI that indicates its property class. 

\subsubsection{RDF Graph Schema.}
RDF Schema (RDFS) \cite{rdfschema} defines a standard vocabulary (i.e., a set of terms, each having a well-defined meaning) which enables the description of resource classes and property classes. From a database perspective, RDF Schema can be used to describe the structure of the data in an RDF database. 

In order to describe classes of resources and properties, the RDF Schema vocabulary defines the following terms:
\code{rdfs:Class} and \code{rdf:Property} represent the classes of resources, and properties respectively; 
\code{rdf:type} can be used (as property) to state that a resource is an instance of a class; 
\code{rdfs:domain} and \code{rdfs:range} allow to define the domain and range of a property, respectively.
Note that \code{rdf:} and \code{rdfs:} are the prefixes for RDF and RDFS respectively.

An RDF Schema is described using RDF triples, so it can be encoded using RDF data formats. The following example shows an RDF Schema which describes the structure of the data shown in Example~\ref{ex:rdfg1}, using the Turtle data format.  
\smallskip
\begin{example}
\label{ex:rdfs1}
\begin{lstlisting}
@prefix rdf: <http://www.w3.org/1999/02/22-rdf-syntax-ns#> .
@prefix xsd: <http://www.w3.org/2001/XMLSchema#>
@prefix voc: <http://www.example.org/voc/> .
voc:Organisation rdf:type rdfs:Class .
voc:name rdf:type rdf:Property ;
         rdfs:domain voc:Organisation ;
         rdfs:range xsd:string .
voc:creation rdf:type rdf:Property ;
             rdfs:domain voc:Organisation ;
             rdfs:range xsd:date .
voc:Person rdf:type rdfs:Class .
voc:birthName rdf:type rdf:Property ;
              rdfs:domain voc:Person ;
              rdfs:range xsd:string .
voc:age rdf:type rdf:Property ;
        rdfs:domain voc:Person ;
        rdfs:range xsd:int .
voc:ceo rdf:type rdf:Property ;
        rdfs:domain voc:Organisation ;
        rdfs:range voc:Person .
\end{lstlisting}
\end{example}
 
Note that a resource class $c$ is defined by a triple of the form \code{($c$ rdf:type rdfs:Class)}.
A property class $pc$ is defined (ideally) by three triples of the form \code{($pc$ rdf:type rdf:Property)}, \code{($pc$ rdfs:domain $rc_1$)} and \code{($pc$ rdfs:range $rc_2$)}, where $rc_1$ indicates the resource class having property  $p$ (i.e. the domain), and $rc_2$ indicates the resource class determining the value of the property $pc$ (i.e. the range). 

If the range of a property class $pc$ is a resource class (defined by the user), then $pc$ is called an object property (e.g. \code{voc:birthName}). 
If the range is a datatype class (defined by RDF Schema or another vocabulary), then $pc$ is called a datatype property.
The IRIs \code{xsd:string}, \code{xsd:integer} and \code{xsd:dateTime} are examples of datatypes defined by XML Schema~\cite{biron2012}.
Let $I_{DT} \subset \dI$ be the set of RDF datatypes.

Note that the RDF schema presented in Example~\ref{ex:rdfs1} provides a complete description of resource classes and property classes.
However, in practice, it is possible to find incomplete or partial RDF schema descriptions. In particular, a property could not define its domain or its range. 

We will assume that a partial schema can be transformed into a total schema. 
In this sense, we will use the term \code{rdfs:Resource}\footnote{According to the RDF Schema specification \cite{rdfschema}, \code{rdfs:Resource} denotes the class of everything.} to complete the definition of properties without domain or range.
For instance, suppose that our sample RDF Schema does not define the range of the property class \code{voc:ceo}. In this case, we include the triple (\code{voc:ceo}, \code{rdfs:range}, \code{rdfs:Resource}) to complete the definition of \code{voc:ceo}. 

Now, we introduce the notion of RDF Graph Schema as a formal way to represent an RDF schema description.
Assume that $I_V \subset \dI$ is a set that includes the RDF Schema terms \code{rdf:type}, \code{rdfs:Class}, \code{rdfs:Property}, \code{rdfs:domain} and \code{rdfs:range}.


\smallskip

\begin{definition}[RDF Graph Schema]\label{def:rdf_schema}
An RDF graph schema is defined as a tuple $S^R = (N_{S}, E_{S}, \phi, \varphi)$ where:
\begin{itemize}
\item $N_S$ is a finite set of nodes representing resource classes; 
\item $E_S$ is a finite set of edges representing property classes;
\item $\phi : (N_S \cup E_S) \to I \setminus I_V$ is a total function that associates each node or edge with an IRI representing a class identifier;
\item $\varphi$: $E_{S} \rightarrow (N_{S} \times N_{S})$ is a total function that associates each property class with a pair of resource classes.
\end{itemize} 
\end{definition}

\smallskip

Recall that $I_{DT}$ denotes the set of RDF datatypes.
Given an RDF Schema description $D$, the procedure to create and RDF Graph schema $S^R = (N_S, E_S, \phi, \varphi)$ from $D$ is given as follows:
\begin{enumerate}
 \item Let $C = \{ rc \mid (rc, \code{rdf:type}, \code{rdfs:Class}) \in D \lor
                           (pc, \code{rdfs:domain},rc) \in D \lor 
                           (pc, \code{rdfs:range},rc) \in D \}$
 \item For each $rc \in C$, we create $n \in N_S$ with $\phi(n) = rc$ 
 \item For each pair of triples
$(pc, \code{rdfs:domain}, rc_1)$ and 
$(pc, \code{rdfs:range}, rc_2)$ in $D$, 
we create $e \in E_S$ with $\phi(e) = pc$ and $\varphi(e) = (n_1,n_2)$, satisfying that $n_1, n_2 \in N_S$, $\phi(n_1) = rc_1$ and $\phi(n_2) = rc_2$. 
\end{enumerate}



Following the above procedure, the RDF schema shown in Example~\ref{ex:rdfs1} can be formally described as follows: 

\smallskip
\begin{lstlisting}
$N_{S} = \{ n_1, n_2, n_3, n_4, n_5 \}$, 
$E_{S} = \{ e_{1}, e_{2}, e_{3}, e_{4}, e_{5} \}$; 
$\phi(n_1) = \{ \text{voc:Organisation} \}$, $\phi(n_2) = \{ \text{voc:Person} \}$, $\phi(n_3) = \{ \text{xsd:date} \}$, $\phi(n_4) = \{ \text{xsd:string} \}$, $\phi(n_5) = \{ \text{xsd:int} \}$ 
$\phi(e_1) = \{ \text{voc:ceo} \}$,  $\phi(e_2) = \{ \text{voc:creation} \}$, $\phi(e_3) = \{ \text{voc:name} \}$, $\phi(e_4) = \{ \text{voc:birthName} \}$, $\phi(e_5) = \{ \text{voc:age} \}$,
$\varphi(e_1) = (n_1, n_2) $, $\varphi(e_2) = (n_1, n_3) $, $\varphi(e_3) = (n_1, n_4) $, $\varphi(e_4) = (n_2, n_4) $, $\varphi(e_5) = (n_2, n_5) $. 
\end{lstlisting} 
\smallskip

\begin{figure}[ht]
    \centering
    \includegraphics[width=\linewidth]{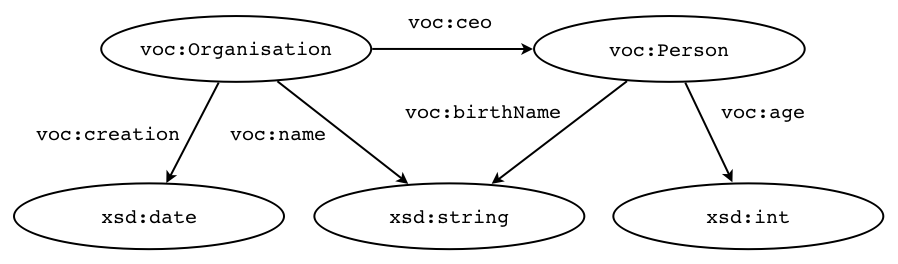}
    \caption{Graphical illustration of an RDF graph schema.}
    \label{fig:rdfs1}
\end{figure}

Additionally, Figure~\ref{fig:rdfs1} shows a graphical representation of the RDF schema graph described above.

Given an RDF graph schema 
$S^R = (N_{S}, E_{S}, \phi, \phi, \varphi)$
and an RDF graph 
$G^R = (N_I, N_L, E_O, E_D, \alpha_I, \alpha_L, \beta_O, \beta_D, \delta)$,
we say that $G^R$ is valid with respect to $S^R$, denoted as $G^R \models S^R$, iff:
\begin{enumerate}
\item for each $r \in N_I \cup N_L$, it applies that there is $rc \in N_S$ where $\delta(r) = \phi(rc)$;
\item for each $e \in E_O$ with $\beta_O(e) = (n,n')$, it applies that there is $pc \in E_S$ where $\delta(e) = \phi(pc)$, $\varphi(pc) = (rc,rc')$, $\delta(n) = \phi(rc)$ and $\delta(n') = \phi(rc')$.
\item for each $e \in E_D$ with $\beta_D(e) = (n,n')$, it applies that there is $pc \in E_S$ where $\delta(e) = \phi(pc)$, $\varphi(pc) = (rc,rc')$, $\delta(n) = \phi(rc)$ and $\delta(n') = \phi(rc')$.
\end{enumerate}

Here, 
condition (1) validates that every resource node is labeled with a resource class defined by the schema; 
condition (2) verifies that each object property edge, and the pairs of resource nodes that it connects, are labeled with the corresponding resource classes; and 
condition (3) verifies that each datatype property edge, and the pairs of nodes that it connects (i.e. a resource node and a literal node), are labeled with the corresponding resource classes

Finally, we present the notion of RDF database.

\smallskip

\begin{definition}[RDF Database]
An RDF database is a pair $(S^R,G^R)$ where $S^R$ is an RDF graph schema and $G^R$ is an RDF graph satisfying that $G^R \models S^R$.   
\end{definition}

\subsection{Property Graph Databases}
\label{sec:pgdm}
A Property Graph (PG) is a labeled directed multigraph whose main characteristic is that nodes and edges can contain a set (possibly empty) of \textit{name-value} pairs referred to as \textit{properties}.
From the point of view of data modeling, each node represents an entity, each edge represents a relationship (between two entities), and each property represents a specific characteristic (of an entity or a relationship).

\begin{figure}[ht]
    \centering
    \includegraphics[width=\linewidth]{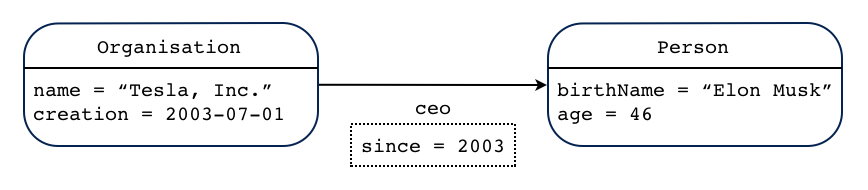}
    \caption{Graphical illustration of a Property Graph.}
    \label{fig:pg1}
\end{figure}

Figure~\ref{fig:pg1} presents a graphical representation of a Property Graph.
The circles represent nodes, the arrows represent edges, and the boxes contain the properties for nodes and edges. 

Currently, there are no standard definitions for the notions of  Property Graph and Property Graph Schema.
However, we present formal definitions that resemble most of the features supported by current PG database systems.

\subsubsection{Property Graph.}
Assume that $\dLab$ is an infinite set of labels (for nodes, edges and properties),  
$\dV$ is an infinite set of (atomic or complex) values, 
and
$\dT$ is a finite set of data types (e.g. \emph{string}, \emph{integer}, \emph{date}, etc.).
A value in $\dV$ will be distinguished as a quoted string.
Given a value $v \in \dV$, the function $\type(v)$ returns the datatype of $v$.
Given a set $S$, $\powerset^+(S)$ denotes the set of non-empty subsets of $S$.

\smallskip

\begin{definition}[Property Graph]
\label{def:pg}
A Property Graph is defined as a tuple
$G^P = (\dN, \dE, \dP, \Gamma, \Upsilon, \Sigma, \Delta)$ where: 
\begin{itemize}
\item $\dN$ is a finite set of nodes, $\dE$ is a finite set of edges, $\dP$ is a finite set of properties, and $\dN, \dE, \dP$ are mutually disjoint sets; 
\item $\Gamma: (\dN \cup \dE) \to \dLab$ is a total function that associates each node or edge with a label;
\item $\Upsilon: \dP \to (\dLab \times \dV )$ is a total function that assigns a label-value pair to each property.
\item $\Sigma: \dE \to (\dN \times \dN)$ is a total function that associates each edge with a pair of nodes; 
\item $\Delta: (\dN \cup \dE) \to \powerset^+(\dP)$ is a partial function that associates a node or edge with a non-empty set of properties, satisfying that $\Delta(o_{1}) \cap \Delta(o_{2}) = \emptyset$ for each pair of objects $o_1, o_2 \in \dom(\Delta)$;
\end{itemize} 
\end{definition}

\smallskip

The above definition supports Property Graphs with the following features:
a pair of nodes can have zero or more edges;
each node or edge has a single label;
each node or edge can have zero or more properties; 
and a node or edge can have the same label-value pair one or more times.

On the other side, the above definition does not support multiple labels for nodes or edges. We have two reasons to justify this restriction. 
First, this feature is not supported by all graph database systems.
Second, it makes complex the definition of schema-instance consistency. 


Given two nodes $n_1, n_2 \in N$ and an edge $e \in E$, satisfying that $\Sigma(e) = (n_1,n_2)$, we will use $e = (n_1,n_2)$ as a shorthand representation for $e$, where $n_1$ and $n_2$ are called the ``source node'' and the ``target node'' of $e$ respectively.

Hence, the formal description of the Property Graph presented in Figure~\ref{fig:pg1} is given as follows:

\smallskip
\begin{lstlisting}
$\dN = \{ n_1, n_2 \}$, 
$\dE = \{ e_1 \}$, 
$\dP = \{ p_1, p_2, p_3, p_4, p_5 \}$, 
$\Gamma(n_1) =  \{\text{Organisation}\}$, $\Gamma(n_2) =  \{\text{Person}\}$, 
$\Gamma(e_1) = \{\text{ceo}\}$, 
$\Upsilon(p_1) = (\text{name}, \text{"Tesla, Inc."})$, $\Upsilon(p_2) = (\text{creation}, \text{2003-07-01})$, $\Upsilon(p_3) = (\text{birthName}, \text{"Elon Musk"})$, $\Upsilon(p_4) = (\text{age}, \text{46})$, $\Upsilon(p_5) = (\text{since}, \text{2003})$
$\Sigma(e_1) = \{ n_1, n_2 \}$, 
$\Delta(n_1) =  \{ p_1, p_2 \}$, $\Delta(n_{2}) =  \{ p_3, p_4 \}$, $\Delta(e_1) =  \{ p_5 \}$.  
\end{lstlisting}
\smallskip

\subsubsection{Property Graph Schema.}\label{sec:pgraph_schema}
A Property Graph Schema defines the structure of a PG database. 
Specifically, it defines types of nodes, types of edges, and the properties for such types.

\begin{figure}[ht]
    \centering
    \includegraphics[width=\linewidth]{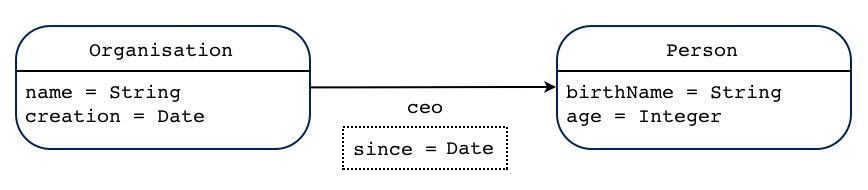}
    \caption{Graphical illustration of a Property Graph Schema.}
    \label{fig:pgs1}
\end{figure}

For instance, Figure~\ref{fig:pgs1} shows a graphical representation of a PG schema.
The formal definition of PG schema is presented next.

\smallskip

\begin{definition}[Property Graph Schema]
\label{def:pg_schema}
A Property Graph Schema is defined as a tuple 
$S^P = (\dN_S, \dE_S, \dP_S, \Theta, \Pi, \Phi, \Psi)$ where:
\begin{itemize}
\item $\dN_S$ is a finite set of node types;
\item $\dE_S$ is a finite set of edge types;
\item $\dP_S$ is a finite set of property types;
\item $\Theta : (\dN_S \cup \dE_S) \to \dLab$ is a total function that assigns a label to each node or edge;
\item $\Pi : \dP_S \to (\dLab \times \dT)$ is a total function that associates each property type with a property label and a data type;
\item $\Phi : \dE_S \to (\dN_S \times \dN_S)$ is a total function that associates each edge type with a pair of node types; 
\item $\Psi : (\dN_S \cup \dE_S) \to \powerset^+(\dP_S)$ is a partial function that associates a node or edge type with a non-empty set of property types, satisfying that $\Psi(o_1) \cap \Psi(o_2) = \emptyset$, for each pair of objects $o_1, o_2 \in \dom(\Psi)$.  
\end{itemize}
\end{definition}

\smallskip

Hence, the formal description of the Property Graph Schema shown in Figure~\ref{fig:pgs1} is the following: 

\smallskip
\begin{lstlisting}
$\dN_S = \{ n_1, n_2 \}$, 
$\dE_S = \{ e_1 \}$, 
$\dP_S = \{ p_1, p_2, p_3, p_4, p_5 \}$,
$\Theta(n_1) = \{ \text{Organisation} \}$, $\Theta(n_2) = \{ \text{Person} \}$, $\Theta(e_1) = \{ \text{ceo} \}$, 
$\Pi(p_1) = (\text{name}, \text{String})$, $\Pi(p_2) = (\text{creation}, \text{Date})$, $\Pi(p_3) = (\text{birthName}, \text{String})$, $\Pi(p_4) = (\text{age}, \text{Integer})$, $\Pi(p_5) = (\text{since}, \text{Date})$, 
$\Phi(e_1) = (n_1, n_2)$,
$\Psi(n_1) =  \{ p_1, p_2 \}$, $\Psi(n_2) =  \{ p_3, p_4 \}$, $\Psi(e_1) =  \{ p_5 \}$ 
\end{lstlisting}
\smallskip

Given a PG schema 
$S^P = (\dN_S, \dE_S, \dP_S, \Theta, \Pi, \Phi, \Psi)$
and a PG 
$G^P = (\dN, \dE, \dP, \Gamma, \Upsilon, \Sigma, \Delta)$,
we say that $G^P$ is valid with respect to $S^P$, denoted as $G^P \models S^P$, iff:
\begin{enumerate}
\item for each $n \in \dN$, it applies that there is $nt \in \dN_S$ satisfying that:
\begin{itemize}
 \item[(a)] $\Gamma(n) = \Theta(nt)$; 
 \item[(b)] for each $p \in \Delta(n)$, there is $pt \in \Psi(nt)$ satisfying that $\Upsilon(p) = (l,v)$ and $\Pi(t_p) = (l, \type(v))$.
\end{itemize}
\item for each $e = (n,n') \in \dE$, it applies that there is $et \in \dE_S$ with $\Phi(et) = (nt,nt')$ satisfying that:
\begin{itemize}
 \item[(a)] $\Gamma(e) = \Theta(et)$, $\Gamma(n) = \Theta(nt)$, $\Gamma(n') = \Theta(nt')$;
 \item[(b)] for each $p' \in \Delta(e)$, there is $pt' \in \Psi(et)$ satisfying that $\Upsilon(p') = (l',v')$ and $\Pi(pt') = (l', \type(v'))$. 
\end{itemize}
\end{enumerate}

Here, condition (1a) validates that every node is labeled with a node type defined by the schema; condition (1b) verifies that each node contains the properties defined by its node type; condition (2a) verifies that each edge, and the pairs of nodes that it connects, are labeled with an edge type, and the corresponding node types; and condition (2b) verifies that each edge contains the properties defined by the schema.

Finally, we present the notion of the Property Graph database.

\smallskip

\begin{definition}[Property Graph Database]
A Property Graph database $D^P$ is a pair $(S^P,G^P)$ where $S^P$ is a PG schema and $G^P$ is a PG satisfying that $G^P \models S^P$.   
\end{definition}

\subsection{RDF databases versus PG databases}
\label{sec:versus}

Upon comparison of RDF graphs and PGs, we see that both share the main characteristics of a traditional labeled directed graph, that is, nodes and edges contain labels, the edges are directed, and multiple edges are possible between a given pair of nodes.
However, there are also some differences between them:
\begin{itemize}
\item An RDF graph contains nodes of type resource (whose label is an IRI) and nodes of type Literal (whose label is a value), whereas a PG allows a single type of node; 
\item Each node or edge in an RDF graph contains just a single value (i.e. a label), whereas each node or edge in a PG could contain multiple labels and  properties respectively;
\item An RDF graph supports multi-value properties, whereas a PG usually just support mono-value properties;
\item An RDF graph allows to have edges between edges, a feature which isn't supported in a PG (by definition);
\item A node in an RDF graph could be associated with zero or more classes or resources, while a node in a PG usually has a single node type.
\end{itemize}


We consider factors such as the availability of schema information in the source model while developing database mappings.
Depending on whether or not the input RDF data has schema, the database mappings can be classified into two types: 
\textbf{(i)} \textit{schema-dependent}: one that generates a target PG schema from the input RDF graph schema, and then transforms the RDF graph into a PG (see Section~\ref{sec:dm1}); and \textbf{(ii)} \textit{schema-independent}: one that creates a generic PG schema (based on predefined structure) and then transforms the RDF graph into a PG (see Section~\ref{sec:dm2}).
In this paper, we developed these two types of database mappings.

Our research omits two special features of RDF: Blank Nodes and reification (i.e. the description of RDF statements using a specific vocabulary). After an empirical study of different RDF datasets e.g. Bio2RDF~\cite{belleau2008}, Europeana~\cite{haslhofer2011}, LOD Cache~\cite{schmachtenberg2014}, Wikidata~\cite{vrandevcic2012}, Billion Triple Challenge (BTC)~\cite{herrera2019} we noticed that these features are rarely used in RDF. Table~\ref{datasets} resume our analysis of the use of Blank Nodes and reification in the above datasets.

\begin{table}[t!]
\centering
\caption{{Blank Nodes and reification in different datasets. Snapshot: 2019-10-08}}
\label{datasets}
\resizebox{0.6\textwidth}{!}{
\begin{tabular}{lll}
\toprule
\textbf{Datasets} & \textbf{Blank Nodes} & \textbf{Reification} \\
\midrule
Europeana         & 0\%                  & 0\%                  \\
Bio2RDF           & 0\%                  & 0\%                  \\
LOD Cache         & 2.67\%               & 1.3\%               \\
Wikidata          & 0.01\%               & 0\%                  \\
BTC               & 12.08\%              & 0.02\%               \\
\bottomrule
\end{tabular}
}
\end{table}

\section{Schema-dependent Database Mapping}
\label{sec:dm1}
In this section we present a database mapping from RDF databases to PG databases.
Specifically, the database mapping $\dm_1$ is composed by the schema mapping $\sm_1$ and the instance mapping $\im_1$.

Recall that $I_{DT}$ is the set of RDF datatypes and $\dT$ is the set of PG datatypes.
Assume that there is a total function $f : I_{DT} \to \dT$ which maps RDF datatypes into PG datatypes.
Additionally, assume that $f^{-1}$ is the inverse function of $f$, i.e. $f^{-1}$ maps PG datatypes into RDF datatypes. 
 
\subsection{Schema mapping \texorpdfstring{$\sm_1$}{Sm}}
We define a schema mapping $\sm_1$ which takes an RDF graph schema as input and returns a PG Schema as output.

\smallskip
\begin{definition}[Schema Mapping $\sm_1$]
\label{def:sm1}
Let 
$S^R = (N_S, E_S, \phi, \varphi)$ be an RDF schema and
$S^P = (\dN_S, \dE_S, \dP_S, \Theta, \Pi, \Phi, \Psi)$ be a Property Graph Schema.
The schema mapping $\sm_1(S^R) = S^P$ is defined as follows:
\begin{enumerate}
 \item For each $rc \in N_S$ satisfying that $\phi(rc) \notin I_C$
  \begin{itemize}
   \item There will be $nt \in \dN_S$ with $\Theta(nt) = \phi(rc)$  
  \end{itemize}
 \item For each $pc \in E_S$ satisfying that $\varphi(pc) = (rc_1,rc_2)$
  \begin{itemize}
   \item If $\phi(rc_2) \in I_{DT}$ then 
    \begin{itemize}
     \item There will be $pt \in \dP_S$ with 
           $\Pi(pt) = (\phi(pc), f(\phi(rc_2)))$,
           $\Psi(nt) = \Psi(nt) \cup pt$ where 
           $nt \in \dN_S$ corresponds to $rc_1 \in N_S$. 
    \end{itemize}
   \item If $\phi(rc_2) \notin I_{DT}$ then 
    \begin{itemize}
     \item There will be $et \in \dE_S$ with
           $\Theta(et) = \phi(pc)$,
           $\Phi(et) = (nt_1,nt_2)$ where $nt_1, nt_2 \in \dN_S$ 
           correspond to $rc_1, rc_2 \in N_S$ respectively.  
    \end{itemize} 
 \end{itemize}  
\end{enumerate}
\end{definition} 
\smallskip

Hence, the schema mapping $\sm_1$ creates a node type for each resource type (with exception of RDF data types); creates a property type for each object property; and creates an edge type for each value property. 

For instance, the Property Graph Schema obtained from the graph schema shown in Figure~\ref{fig:rdfs1} is given as follows:

\smallskip
\begin{lstlisting}
$\dN_S = \{ n_1, n_2 \}$, 
$\dE_S = \{ e_1 \}$, 
$\dP_S = \{ p_1, p_2, p_3, p_4 \}$,
$\Theta(n_1) = \{ \text{voc:Organisation} \}$, $\Theta(n_2) = \{ \text{voc:Person} \}$, $\Theta(e_1) = \{ \text{voc:ceo} \}$, 
$\Pi(p_1) = (\text{voc:creation}, \text{Date})$, $\Pi(p_2) = (\text{voc:name}, \text{String})$, $\Pi(p_3) = (\text{voc:birthName}, \text{String})$, $\Pi(p_4) = (\text{voc:age}, \text{Integer})$,
$\Phi(e_1) = (n_1, n_2)$,
$\Psi(n_1) =  \{ p_1, p_2 \}$, $\Psi(n_2) =  \{ p_3, p_4 \}$. 
\end{lstlisting}
\smallskip

\subsection{Instance Mapping \texorpdfstring{$\im_1$}{Im1}}
Now, we define the instance mapping $\im_1$ which takes an RDF graph as input and returns a Property Graph as output.

\smallskip
\begin{definition}[Instance Mapping $\im_1$]
\label{def:im1}
Let --\\
$G^R = (N_I, N_L, E_O, E_D, \alpha_I, \alpha_L, \beta_O, \beta_D, \delta)$ be an RDF graph and 
$G^P = (\dN, \dE, \dP, \Gamma, \Upsilon, \Sigma, \Delta)$ be a Property Graph.
The instance mapping $\im_1(G^R) = G^P$ is defined as follows:
\begin{enumerate}
\item For each $r \in N_I$
 \begin{itemize}
  \item There will be $n \in \dN$ with $\Gamma(n) = \delta(r)$.
  \item There will be $p \in \dP$ with 
        $\Upsilon(p) = (\text{iri},\alpha_I(r))$
  \item $\Delta(n) = \{ p \}$.
 \end{itemize}
\item For each $dp \in E_D$ satisfying that $\beta_D(dp) = (r_1,r_2)$
 \begin{itemize}
  \item There will be $p \in \dP$ with 
        $\Upsilon(p) = (\delta(dp), \alpha_L(r_2))$,
        $\Delta(n) = \Delta(n) \cup \{ p \}$  where
        $n \in \dN$ corresponds to $r_1 \in N_I$. 
 \end{itemize}
\item For each $op \in E_O$ satisfying that $\beta_O(op) = (r_1,r_2)$
 \begin{itemize}
  \item There will be $e \in \dE$ with
        $\Gamma(e) = \delta(op)$,
        $\Sigma(e) = (n_1,n_2)$ where 
        $n1,n2 \in \dN$ correspond to $r1,r2 \in N_I$ respectively. 
 \end{itemize} 
\end{enumerate}
\end{definition} 
\smallskip

According to the above definition, the instance mapping $\im_1$ creates a node in $G_R$ for each resource node, creates a property in $G_R$ for each datatype property, and creates an edge in $G_R$ for each object property. 

For example, the PG obtained from the RDF graph is shown in Figure~\ref{fig:rdfg1} is given as follows:

\smallskip
\begin{lstlisting}
$\dN = \{ n_1, n_2 \}$, 
$\dE = \{ e_1 \}$, 
$\dP = \{ p_1, p_2, p_3, p_4, p_5, p_6 \}$, 
$\Gamma(n_1) =  \{\text{voc:Organisation}\}$, $\Gamma(n_2) =  \{\text{voc:Person}\}$, 
$\Gamma(e_1) = \{\text{voc:ceo}\}$, 
$\Upsilon(p_1) = (\text{iri}, \text{"ex:Test\_Inc"})$, $\Upsilon(p_2) = (\text{voc:name}, \text{"Tesla, Inc."})$, $\Upsilon(p_3) = (\text{voc:creation}, \text{2003-07-01})$, $\Upsilon(p_4) = (\text{iri}, \text{"ex:Elon\_Musk"})$ $\Upsilon(p_5) = (\text{voc:birthName}, \text{"Elon Musk"})$, $\Upsilon(p_6) = (\text{voc:age}, \text{46})$,
$\Sigma(e_1) = \{ n_1, n_2 \}$, 
$\Delta(n_1) =  \{ p_1, p_2, p_3 \}$, $\Delta(n_{2}) =  \{ p_4, p_5, p_6 \}$.  
\end{lstlisting}
\smallskip

\subsection{Properties of \texorpdfstring{$\dm_1$}{Dm1}}
\label{sec:propdm1}
In this section, the database mapping $\dm_1 = (\sm_1,\im_1)$ will be evaluated with respect to the properties described in Section~\ref{sec:properties}. 
Specifically, we will analyze computability, semantics preservation, and information preservation.

Recall that $\dm_1$ is a formed by the schema mapping $\sm_1$ and the instance mapping $\im_1$.

\smallskip
\begin{proposition}
The database mapping $\dm_1$ is computable.
\end{proposition}
\smallskip

It is straightforward to see that Definition~\ref{def:sm1} and Definition~\ref{def:im1} can be transformed into algorithms to compute $\sm_1$ and $\im_1$ respectively.

\smallskip
\begin{lemma}
The database mapping $\dm_1$ is \emph{semantics preserving}.
\end{lemma}
\smallskip

Note that the schema mapping $\sm_1$ and the instance mapping $\im_1$ have been designed to create a Property Graph database that maintains the restrictions defined by the source RDF database.
On one side, the schema mapping $\sm_1$ allows transforming the structural and semantic restrictions from the RDF graph schema to the PG schema.
On the other side, any Property Graph generated by the instance mapping will be valid with respect to the generated PG schema.   

The semantics preservation property of $\dm_1$ is supported by the following facts:
\begin{itemize}
\item We provide a procedure to create a complete RDF graph schema $S^R$ from a set of RDF triples describing an RDF schema, i.e. each property defines its domain and range resource classes. 
\item We provide a procedure to create an RDF graph $G^R$ from a set of RDF triples, satisfying that each every node and edge in $G^R$ is associated with a resource class; it allows a complete connection between the RDF instance and the RDF schema. 
\item The schema mapping $\sm_1$ creates a node type for each user-defined resource type, a property type for each datatype property edge, and an edge for each object property type.
\item Similarly, the instance mapping $\im_1$ creates a node for each resource, a property for each resource-literal edge, and an edge for each resource-resource edge.
\end{itemize}

\smallskip
\begin{theorem}
The database mapping $\dm_1$ is \emph{information preserving}.
\end{theorem}
\smallskip

In order to prove that $\dm_1$ is information preserving, we need to define a database mapping $\dm^{-1}_1 = (\sm^{-1}_1, \im^{-1}_1)$ which allows to transform a PG database into an RDF database, and satisfying that $D = \dm^{-1}_1(\dm_1(D))$ for any RDF database $D$.
Next we define the schema mapping $\sm^{-1}_1$ and the instance mapping $\im^{-1}_1$.   

\smallskip
\begin{definition}[Schema mapping $\sm^{-1}_1$]
Let 
$S^P = (\dN_S, \dE_S, \dP_S, \Theta, \Pi, \Phi, \Psi)$ be a Property Graph Schema and 
$S^R = (N_S, E_S, \phi, \varphi)$ be an RDF schema.
The schema mapping $\sm^{-1}_1(S^R) = S^P$ is defined as follows:
\begin{enumerate}
\item Let $T = \{ dt \mid (pl,dt) \in \range(\Pi) \}$ 
\item For each $dt \in T$, we create $rc \in N_S$ with $\phi(rc) = f^{-1}(dt)$
\item For each $et \in \dE$ with $\Phi(et) = (nt_1, nt_2)$, 
      we create $pc \in E_S$ with $\phi(pc) = \Theta(et)$
      and $\varphi(pc) = (rc_1, rc_2)$ satisfying that
      $rc_1, rc_2 \in N_S$, $\phi(rc_1) = \Theta(nt_1)$ 
      and $\phi(rc_2) = \Theta(nt_2)$
\item For each $nt \in \dN_S$, we create $rc \in N_S$ with $\phi(rc) = \Theta(nt)$
 \begin{enumerate}
   \item For each $pt \in \Psi(nt)$ such $\Pi(pt)=(pl,dt)$, 
         we create $pc \in E_S$ with $\phi(pc) = pl$ 
         and $\varphi(pc) = (rc_1, rc_2)$ satisfying that
         $rc_1, rc_2 \in N_S$, $\phi(rc_1) = \Theta(nt)$ 
         and $\phi(rc_2) = f^{-1}(dt)$  
   \begin{enumerate}
    \item There will be $pc \in E_S$ with \
   \end{enumerate} 
 \end{enumerate}
\end{enumerate}
\end{definition}
\smallskip

In general terms, the schema mapping $\sm^{-1}_1$ creates a resource class for each node type, an object property for each edge type, and a datatype property for each property type. 
Given a PG schema $S^P = \sm_1(S^R)$, the schema mapping $\sm^{-1}_1$ allows to ``recover'' all the schema constraints defined by $S^R$, i.e $\sm^{-1}_1(S^P) = S^R$.  

An issue of $\sm^{-1}_1$, is the existence of RDF datatypes which are not supported by PG databases. For example, \code{rdfs:Literal} has no equivalent datatype in PG database systems. The solution to this issue is to find a one-to-one correspondence between RDF datatypes and PG datatypes. 

\smallskip
\begin{definition}[Instance mapping $\im^{-1}_1$]
Let $G^P = (\dN, \dE, \dP, \Gamma, \Upsilon, \Sigma, \Delta)$ be a Property Graph and $G^R = (N_I, N_L, E_O, E_D, \alpha_I, \alpha_L, \beta_O, \beta_D, \delta)$ be an RDF graph.
The instance mapping $\im^{-1}_1(G^P) = G^R$ is defined as follows:
\begin{enumerate}
 \item For each $n \in \dN$, there will be $r \in N_I$ where
 \begin{enumerate}
  \item $\alpha_I(r) = v$ such that $p \in \Delta(n)$ 
        and $\Upsilon(p) = (\code{iri}, v)$
  \item $\delta(r) = \Gamma(n)$
  \item For each $p \in \Delta(n) \setminus \{\code{iri}\}$ 
        where $\Upsilon(p) = (lab,val)$,
        there will be $l \in N_L$ and $dp \in E_D$ with
        $\alpha_L(l) = val$, 
        $\delta(l) = f^{-1}(\type(val))$, 
        $\delta(dp) = lab$ and
        $\beta(dp) = (r,l)$
 \end{enumerate}
 \item For each $e \in \dE$ where $\Sigma(e) = (n_1,n_2)$, 
       there will be $op \in E_O$ with 
       $\delta(op) = \Gamma(e)$ and
       $\beta(op) = (r_1,r_2)$ such that $r_1,r_2$ 
       correspond to $n_1,n_2$ respectively.
\end{enumerate}
\end{definition}
\smallskip

Hence, the method defined above defines that each node in $G^P$ is transformed into a resource node in $G^R$, each property in $G^P$ is transformed into a datatype property in $G^R$, and each edge in $G^P$ is transformed into an object property in $G^R$.
Given a Property Graph $G^P = \im_1(G^R)$, the instance mapping $\im^{-1}_1$ allows to ``recover'' all the data in $G^R$, i.e $\im^{-1}_1(G^P) = G^R$.  

Note that each RDF graph produced by the instance mapping $\im^{-1}_1$ will be valid with respect to the schema produced with the corresponding schema mapping $\sm^{-1}_1$.
Hence, any RDF database $D^R$ can be transformed into a PG database by using the database mapping $\dm(D^R)$, and $D^R$ could be recovered by using the database mapping $\dm^{-1}_1$.

\section{Schema-independent Database Mapping}
\label{sec:dm2}
In this section we present a database mapping, from RDF databases to PG databases, which does not consider the database schema.
In this sense, we provide a method to transform any RDF graph into a Property Graph database.

Given an RDF database $D^R = (S^R,G^R)$, we will define the database mapping $\dm_2 = (\sm_2,\im_2)$ which allows to construct a PG database 
$D^P = (S^*,\im_2(G^R))$ where $\sm_2$ is a schema mapping that, for any RDF graph $G^R$, $\sm_2$ creates a generic Property Graph Schema $S^*$.    

\subsection{Generic Property Graph Schema}
First we introduce a Property Graph Schema which is able to model any RDF graph.  

\smallskip
\begin{definition}[Generic Property Graph Schema]
\label{def:gpg}
Let -- \\ $S^* = (\dN_S, \dE_S, \dP_S, \Theta, \Pi, \Phi, \Psi)$ be the Property Graph Schema defined as follows:
\begin{lstlisting}
$\dN_S = \{ n_1, n_2 \}$, 
$\dE_S = \{ e_1, e_2 \}$, 
$\dP_S = \{ p_1, p_2, p_3, p_4, p_5, p_6 \}$,
$\Theta(n_1) = \{ \text{Resource} \}$, $\Theta(n_2) = \{ \text{Literal} \}$, $\Theta(e_1) = \{ \text{ObjectProperty} \}$, $\Theta(e_1) = \{ \text{DatatypeProperty} \}$
$\Pi(p_1) = (\text{iri}, \text{String})$, $\Pi(p_2) = (\text{type}, \text{String})$, $\Pi(p_3) = (\text{value}, \text{String})$, $\Pi(p_4) = (\text{type}, \text{String})$, $\Pi(p_5) = (\text{type}, \text{String})$, $\Pi(p_6) = (\text{type}, \text{String})$
$\Phi(e_1) = (n_1, n_1)$, $\Phi(e_1) = (n_1, n_2)$
$\Psi(n_1) =  \{ p_1, p_2 \}$, $\Psi(n_2) =  \{ p_3, p_4 \}$, $\Psi(e_1) =  \{ p_5 \}$, $\Psi(e_2) =  \{ p_6 \}$ 
\end{lstlisting}
\end{definition} 

\begin{figure}[ht]
    \centering
    \includegraphics[width=\linewidth]{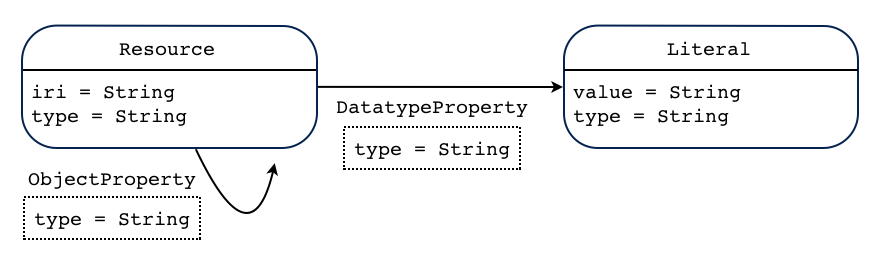}
    \caption{A generic Property Graph Schema.}
    \label{fig:pgsg}
\end{figure}

In the above definition: 
the node type \code{Resource} will be used to represent RDF resources,
the node type \code{Literal} will be used to represent RDF literals,
the edge type \code{ObjectProperty} allows to represent object properties (i.e. relationships between RDF resources), and 
the edge type \code{DatatypeProperty} allows representing datatype properties (i.e. relationships between an RDF resource an a literal).  
Figure~\ref{fig:pgsg} shows a graphical representation of the generic Property Graph Schema. 

\subsection{Instance Mapping \texorpdfstring{$\im_2$}{Im}}
Now, we define the instance mapping $\im_2$ which takes an RDF graph and produces a Property Graph following the restrictions established by the generic Property Graph Schema defined above.

\smallskip
\begin{definition}[Instance mapping $\im_2$]
\label{def:im2}
Let -- \\
$G^R = (N_I, N_L, E_O, E_D, \alpha_I, \alpha_L, \beta_O, \beta_D, \delta)$ be an RDF graph
and 
$G^P = (\dN, \dE, \dP, \Gamma, \Upsilon, \Sigma, \Delta)$
be a Property Graph.
The instance mapping $\im_2(G^R) = G^P$ is defined as follows:
\begin{enumerate}
\item For each $r \in N_I$
 \begin{itemize}
  \item There will be $n \in \dN$ with 
        $\Gamma(n) = \code{Resource}$
  \item There will be $p \in \dP$ with 
        $\Upsilon(p) = (\code{iri},\alpha_I(r))$
  \item There will be $p' \in \dP $ with 
        $\Upsilon(p') = (\code{type},\delta(r))$
  \item $\Delta(n) = \Delta(n) \cup p \cup p'$
 \end{itemize}
\item For each $l \in N_L$
 \begin{itemize}
  \item There will be $n \in \dN$ with 
        $\Gamma(n) = \code{Literal}$
  \item There will be $p \in \dP$ with 
        $\Upsilon(p) = (\code{value},\alpha_L(l))$
  \item There will be $p' \in \dP $ with 
        $\Upsilon(p') = (\code{type},\delta(l))$
  \item $\Delta(n) = \Delta(n) \cup p \cup p'$
 \end{itemize}
\item For each $dp \in E_D$ satisfying that $\beta_D(dp) = (r_1,r_2)$
 \begin{itemize}
  \item There will be $e \in \dE$ with 
        $\Gamma(e) = \code{DatatypeProperty}$, and
        $\Sigma(e) = (n_1,n_2)$ where $n_1,n_2 \in \dN$ 
        correspond to $r_1,r_2 \in N_I$ respectively
  \item There will be $p \in \dP$ with 
        $\Upsilon(p) = (\code{type},\delta(dp))$
  \item $\Delta(e) = \Delta(e) \cup p$
 \end{itemize} 
\item For each $op \in E_O$ satisfying that $\beta_O(op) = (r_1,r_2)$
 \begin{itemize}  
  \item There will be $e \in \dE$ with 
        $\Gamma(e) = \code{ObjectProperty}$, and
        $\Sigma(e) = (n_1,n_2)$ where $n_1,n_2 \in \dN$ 
        correspond to $r_1,r_2 \in N_I$ respectively
  \item There will be $p \in \dP$ with 
        $\Upsilon(p) = (\code{type},\delta(op))$
  \item $\Delta(e) = \Delta(e) \cup p$  
 \end{itemize} 
\end{enumerate}
\end{definition} 

According to the above definition, the instance mapping $\im_2$ creates 
PG nodes from resource nodes and literal nodes, and
PG edges from datatype properties and object properties.
The property \code{type} is used to maintain resource class identifiers and RDF datatypes.
The property \code{iri} is used to store the IRI of RDF resources and properties.
The property \code{value} is used to maintain a literal value.  

For example, the PG obtained after applying $\im_2$ over the RDF graph shown in Figure~\ref{fig:rdfg1} is given as follows:

\smallskip
\begin{lstlisting}
$\dN = \{ n_1, n_2, n_3, n_4, n_5, n_6 \}$, 
$\dE = \{ e_1, e_2, e_3, e_4, e_5 \}$, 
$\dP = \{ p_1, \dots, p_{17} \}$, 
$\Gamma(n_1) =  \{\text{Resource}\}$, $\Gamma(n_2) =  \{\text{Resource}\}$, $\Gamma(n_3) =  \{\text{Literal}\}$, $\Gamma(n_4) =  \{\text{Literal}\}$, $\Gamma(n_5) =  \{\text{Literal}\}$, $\Gamma(n_6) =  \{\text{Literal}\}$, 
$\Gamma(e_1) = \{\text{ObjectProperty}\}$, $\Gamma(e_2) = \{\text{DatatypeProperty}\}$, $\Gamma(e_3) = \{\text{DatatypeProperty}\}$, $\Gamma(e_4) = \{\text{DatatypeProperty}\}$, $\Gamma(e_5) = \{\text{DatatypeProperty}\}$, 
$\Upsilon(p_1) = (\text{iri}, \text{"ex:Tesla\_Inc"})$, $\Upsilon(p_2) = (\text{type}, \text{"voc:Organisation"})$, $\Upsilon(p_3) = (\text{iri}, \text{"ex:Elon\_Musk"})$, $\Upsilon(p_4) = (\text{type}, \text{"voc:Person"})$ $\Upsilon(p_5) = (\text{value}, \text{"Tesla, Inc."})$, $\Upsilon(p_6) = (\text{type}, \text{"xsd:string"})$, $\Upsilon(p_7) = (\text{value}, \text{"2003-07-01"})$, $\Upsilon(p_8) = (\text{type}, \text{"xsd:date"})$, $\Upsilon(p_9) = (\text{value}, \text{"Elon Musk"})$, $\Upsilon(p_{10}) = (\text{type}, \text{"xsd:string"})$, $\Upsilon(p_{11}) = (\text{value}, \text{"46"})$, $\Upsilon(p_{12}) = (\text{type}, \text{"xsd:int"})$, $\Upsilon(p_{13}) = (\text{iri}, \text{"voc:ceo"})$, $\Upsilon(p_{14}) = (\text{iri}, \text{"voc:name"})$, $\Upsilon(p_{15}) = (\text{iri}, \text{"voc:creation"})$, $\Upsilon(p_{16}) = (\text{iri}, \text{"voc:birthName"})$, $\Upsilon(p_{17}) = (\text{iri}, \text{"voc:age"})$
$\Sigma(e_1) = \{ n_1, n_2 \}$, $\Sigma(e_2) = \{ n_1, n_3 \}$, $\Sigma(e_3) = \{ n_1, n_4 \}$, $\Sigma(e_4) = \{ n_2, n_5 \}$, $\Sigma(e_5) = \{ n_2, n_6 \}$, 
$\Delta(n_1) =  \{ p_1, p_2 \}$, $\Delta(n_2) =  \{ p_3, p_4 \}$, $\Delta(n_3) =  \{ p_5, p_6 \}$, $\Delta(n_4) =  \{ p_7, p_8 \}$, $\Delta(n_5) =  \{ p_9, p_{10} \}$, $\Delta(n_6) =  \{ p_{11}, p_{12} \}$, 
$\Delta(e_1) =  \{ p_{13} \}$, $\Delta(e_2) =  \{ p_{14} \}$, $\Delta(e_3) =  \{ p_{15} \}$, $\Delta(e_4) =  \{ p_{16} \}$, $\Delta(e_5) =  \{ p_{17} \}$.  
\end{lstlisting}
\smallskip

\begin{figure}[ht]
    \centering
    \includegraphics[width=\linewidth]{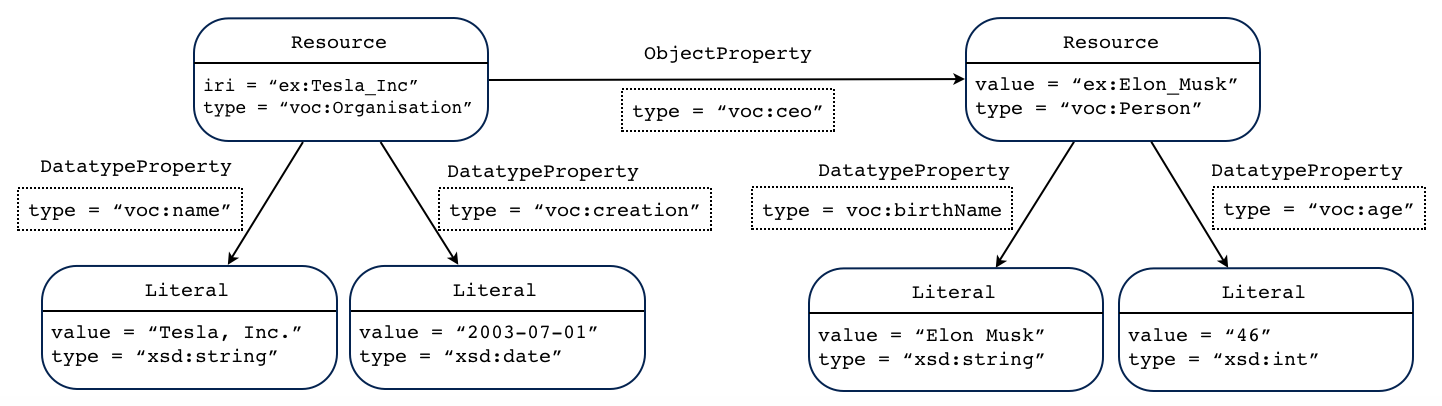}
    \caption{Example of Property Graph obtained after applying the schema-independent database mapping.}
    \label{fig:pg2}
\end{figure}

Figure~\ref{fig:pg2} shows a graphical representation of the Property Graph described above.

\subsection{Properties of \texorpdfstring{$\dm_2$}{Dm}}
\label{sec:propdm2}
In this section, we evaluate the computability, semantics preservation and information preservation of the database mapping $\dm_2$. 

Recall that $\dm_2$ is a formed by the schema mapping $\sm_2$ and the instance mapping $\im_2$, where $\sm_2$ always creates a generic PG schema $S^*$.

\smallskip
\begin{proposition}
The database mapping $\dm_2$ is \emph{computable}.
\end{proposition}
\smallskip

It is not difficult to see that an algorithm can be created from the description of the instance mapping $\im_2$, presented in Definition~\ref{def:im2}.

\smallskip
\begin{lemma}
The database mapping $\dm_2$ is \emph{semantics preserving}.
\end{lemma}
\smallskip

It is straightforward to see (by definition) that any PG graph created with the instance mapping $\im_2$ will be valid with respect to the generic PG schema $S^*$.    

\smallskip
\begin{theorem}
The database mapping $\dm_2$ is \emph{information preserving}.
\end{theorem}
\smallskip

In order to prove that $\dm_2$ is information preserving, we need to provide  a database mapping $\dm^{-1}_2$ which allows to transform a PG database into an RDF database, and show that for every RDF database $D^R$, it applies that $D^R = \dm^{-1}_2( \dm_2 (D^R)$.

Recalling that the objective of this section is to provide a schema-independent database mapping, we will assume that for any RDF database $D^R = (S^R,G^R)$, the RDF graph schema  $S^R$ is null or irrelevant to validate $G^R$.
Hence, we just define an instance mapping $\im^{-1}_2$ which allows to transform a PG graph into an RDF database, such that for every RDF graph $G^R$, it must satisfy that $G^R = \im^{-1}_2 ( \im_2 (G^R) )$. 

\smallskip
\begin{definition}[Instance mapping $\im^{-1}_2$]
Let $G^P = (\dN, \dE, \dP, \Gamma, \Upsilon, \Sigma, \Delta)$ be a Property Graph and $G^R = (N_I, N_L, E_O, E_D, \alpha_I, \alpha_L, \beta_O, \beta_D, \delta)$ be an RDF graph.
The instance mapping $\im^{-1}_2(G^P) = G^R$ is defined as follows:
\begin{enumerate}
\item For each $n \in \dN$ satisfying that $\Gamma(n) = \code{Resource}$, $p_1, p_2 \in \Delta(n)$, $\Upsilon(p_1) = (\code{iri}, v_1)$ and $\Upsilon(p_2) = (\code{type},v_2)$, then there will be $r \in N_I$ with $\alpha_I(r) = v_1$ and $\delta(r) = v_2$ 
\item For each $n \in \dN$ satisfying that $\Gamma(n) = \code{Literal}$, $p_1, p_2 \in \Delta(n)$, $\Upsilon(p_1) = (\code{value}, v_1)$ and $\Upsilon(p_2) = (\code{type},v_2)$, then there will be $r \in N_L$ with $\alpha_L(r) = v_1$ and $\delta(r) = v_2$ 
\item For each $e \in \dE$ satisfying that $\Gamma(e) = \code{ObjectProperty}$ , $p \in \Delta(e)$, $\Upsilon(p) = (\code{type},v)$, $\Sigma(e) = (n_1,n_2)$, then there will be $op \in E_O$ with $\delta(op) = v$, $\beta_O(op) = (r_1, r_2)$ where $r_1 \in N_I$ corresponds to $n_1 \in \dN$, and $r_2 \in N_I$ corresponds to $n_2 \in \dN$ 
\item For each $e \in \dE$ satisfying that $\Gamma(e) = \code{DatatypeProperty}$ , $p \in \Delta(e)$, $\Upsilon(p) = (\code{type},v)$, $\Sigma(e) = (n_1,n_2)$, then there will be $dp \in E_D$ with $\delta(dp) = v$, $\beta_D(dp) = (r_1, r_2)$ where $r_1 \in N_I$ corresponds to $n_1 \in \dN$, and $r_2 \in N_L$ corresponds to $n_2 \in \dN$  
\end{enumerate}
\end{definition}
\smallskip

Hence, the method defined above defines that for each node labeled with \code{Resource} is transformed into a resource node, each node labeled with \code{Literal} is transformed into a literal node, each edge labeled with \code{ObjectProperty} is transformed into a resource-resource edge, and each edge labeled with \code{DatatypeProperty} is transformed into a resource-literal edge. 
Additionally, the property \code{iri} is used to recover the original IRI identifier (for nodes), and the property \code{type} allows us to recover the IRI identifier of the resource class associated to each node or edge. 

It is not difficult to verify that for any RDF graph $G^R$, we can produce a PG graph $G^P = \sm_1(G^R)$, and then recover $G^R$ by using $\im^{-1}_2 ( G^P )$.

\section{Related Work}
\label{sec:relwork}
In this section we present the related work that targets the interoperability issue between the RDF and PG data models. 
We group the efforts based on mapping of the data model they target, i.e. RDF $\rightarrow$ PG and vice versa, and summarise their shortcomings.

On the other hand, how to build RDF from existing data sources becomes an important issue in the area of Semantic Web. For that purpose, many papers come up with transforming methods of constructing RDF from different sources, e.g. XML~\cite{bischof2012,thuy2009}, relational databases~\cite{bizer2004,das2012}. However, there are also more general approaches~ \cite{corby2015a,corby2015b,connolly2007,dimou2014}.

\smallskip
\noindent
\textbf{RDF $\rightarrow$ PG.}
Hartig~\cite{hartig2019foundations} proposes two formal transformations between RDF$^{\star}$ and PGs. 
RDF$^{\star}$ is a conceptual extension of RDF which is based on reification. The first transformation maps any RDF triple as an edge in the resulting PG. Each node has the ``kind'' property to describe the node type (e.g. IRI). 
The second transformation distinguishes data and object properties. The former is transformed into node properties and latter into edges of a PG.    
The shortcoming of this approach is that RDF$^{\star}$ -- (i) does not support mapping an RDF graph schema, and (ii) adds an extra step of an intermediate mapping and; (iii) isn't supported by major RDF stores. 

In \textit{S2X}, Sch{\"a}tzle et al.~\cite{schatzle2015s2x} propose a GraphX-specific RDF-PG transformation. 
The mapping uses attribute \textit{label} to store the node and edge identifiers, i.e. each triple \textit{t = (s, p, o)} is represented using two vertices $v_{s}, v_{0}$, an edge ($v_{s}, v_{o}$) and labels $v_{s}$.\textit{label = s}, $v_{o}$.\textit{label = o}, ($v_{s}, v_{o}$).\textit{label = p}. 
Apart from being only GraphX-specific, this approach misses the concept of properties and also does not cover RDF graph Schema.

Nyugen et al.~\cite{nguyen2016formal}, propose the \textit{LDM-3N} (labeled directed multigraph - three nodes) graph model. 
This data model represents each triple element as separate nodes, thus the three nodes (3N) . 
The \textit{LDM-3N} graph model is used to address the Singleton Property (SP) based reified RDF data. 
The problem with this approach is that: 
(i) it adds adds an extra computation step (and \textit{2n} triples);  
(ii) does not cover RDF graph Schema; and misses the concept of properties.

Tomaszuk~\cite{tomaszuk2016rdf}, propose \textit{YARS} serialization for transforming RDF data into PGs.
This approach performs only a syntactic transformation between encoding schemes and does not cover RDF Schema. 

\smallskip
\noindent
\textbf{PG $\rightarrow$ RDF.}
There exist very few proposals for the PG-to-RDF transformation, such as Das et al.~\cite{das2014tale} and Hartig~\cite{hartig2019foundations}, that mainly use RDF \textit{reification} methods (including Blank Nodes) to convert nodes and edge properties in a PG to RDF data.
While~\cite{hartig2019foundations} propose an in-direct mapping that requires converting to the $RDF^{\star}$ model (as mentioned earlier),  \cite{das2014tale} lacks a formal foundation.
Both approaches do not consider the presence of a PG schema. Another approach is Unified Relational Storage (URS) \cite{zhang2019}. It focuses on interchangeably managing RDF and Property Graphs, and this is not a strict transformation method.

Table~\ref{tab:my-table}, presents a consolidated summary of related work and the features they address.

\begin{table*}[ht]
\centering
\caption{{A consolidated summary of related work addressing data interoperability. Here "-" implies no evidence/claim was found the respective study.}}
\label{tab:my-table}
\resizebox{\textwidth}{!}{
\begin{tabular}{p{2.4cm}p{2cm}p{1.7cm}p{1.5cm}p{1.5cm}p{2.5cm}}
\toprule
{{Work}} & {{Target}} & {{Is a Direct mapping?}} & Formally Defined? & Covers Schema? & Is Information Preserving? 
\\ \midrule
{$\text{RDF}^{\star}$ \cite{hartig2019foundations}} & {RDF$\leftrightarrow$PG} & {No} & {Yes} & {No} & {--} \\
{URS \cite{zhang2019}} & {RDF$\leftrightarrow$PG} & {No} & {No} & {Yes} & {No}   \\
{S2X \cite{schatzle2015s2x}} & {RDF$\rightarrow$PG} & - &  {No} & {No} & {No}   \\
{LDM-3N \cite{nguyen2016formal}} & {RDF$\rightarrow$PG} & {No} &  {No} & {No} & {No} 
\\
{Das et al. \cite{das2014tale}} & {PG$\rightarrow$RDF} & {No} & {No} & {No} & {No}    \\
{YARS \cite{tomaszuk2016rdf}} & {PG$\rightarrow$RDF} & {No} & {No} & {No} & {No}   \\

{{Our approach}} & {RDF$\leftrightarrow$PG} & {Yes} & {Yes} & {Yes} & {Yes}
\\ \bottomrule
\end{tabular}
}
\end{table*}

\section{Conclusions}
\label{sec:conclu}
In this article, we have presented two mappings to transform RDF databases into Property Graph databases. 
We have shown that both database mappings satisfy the property of information preservation, i.e. there exist inverse mappings that allow recovering the original databases without losing information.
These results allow us to present the following conclusion about the information capacity of the Property Graph data models with respect to the RDF data model.

\smallskip
\begin{corollary}
The property graph data model subsumes the information capacity of the RDF data model.
\end{corollary}
\smallskip

Although our methods assume some condition for the input RDF databases, they are generic and can be easily extended (by overloading the mapping functions) to provide support for features such as inheritance (e.g. \code{rdfs:subClassOf}, \code{rdfs:subPropertyOf}) and Blank Nodes (via Skolemisation~\cite{91033}).
Furthermore, our formal definitions will be very useful to study query interoperability~\cite{thakkar2018two,thakkar2018stitch,thakkar2020let,nguyen2019singleton} and query preservation between RDF and Property Graph databases (i.e. query transformation between SPARQL queries and any Property Graph query language).
Thus, with this paper, we take a substantial step by laying the core formal foundation for supporting interoperability between RDF and Property Graph databases. 
As future work, we plan to incorporate features such as RDF reification techniques (N-ary Relations, Named Graphs, etc), inheritance, and OWL 2 RL.


\bibliography{ref}
\bibliographystyle{plain}

\end{document}